# Mining the SPARC Galaxy Database: Finding "Hidden Variables" in the Baryonic Tully-Fisher Relation

Jeffrey M. La Fortune
1081 N. Lake St. Neenah, WI 54956 forch2@gmail.com
6 January 2021

*Abstract*
The Baryonic Tully-Fisher Relation (BTFR) links baryonic mass of rotationally supported galaxies to their flat disk velocities. A popular form of the BTFR linked to MOND is based on an empirically determined characteristic acceleration, $a_0$ that serves as the constant of proportionality. In this work, we propose an alternative, parametric form of the BTFR employing individual galactic properties from the SPARC galaxy database. Based on this data we find that a precise mass-velocity correlation is possible taking into account galactic disk radius and two dynamical related properties; mass discrepancy and disk surface density. We find no need to invoke a characteristic acceleration constant although its ansatz can be extracted and compared to several recent analyses also arguing against the MOND interpretation of $a_0$. This improved BTFR finding has ramifications for the Radial Acceleration Relation (RAR) as well. Rather than a universal relation that describes the dynamics of all rotationally-supported galaxies, we find the RAR consists of a statistically distributed family of curves, reflecting the unique properties attributed to individual galaxies.

*Introduction*
The Baryonic Tully-Fisher Relation (BTFR) relates the baryonic mass $M_{Bar}$ with a galactic rotation velocity $V_f$ (or circular velocity $V_C$) in the flat part of the curve. In the conventional form shown below, G = gravitational constant and $a_0 = g†$ is an empirically derived 'characteristic' acceleration constant. The BTFR is typically expressed as:

$$V_f^4 = a_0 G M_{Bar} \ or \ V_C^4 = g^† G M_{Bar}$$

The velocity exponent of four is derived from dynamic and baryonic acceleration components in the model via $g_{Obs}=(g_{Dyn})=V_C^2/R$ and $g_{Bar}=GM_{Bar}/R^2$, respectively. To maintain correct dimensionality, an acceleration term ($a_0$ or $g†$ as above) is required and is deemed a constant for all rotationally-supported galaxies. The left-hand equation has origins in the theory of MOdified Newtonian Dynamics (MOND) introduced by Milgrom as an alternative to dark matter to solve the galactic 'missing mass' problem (Milgrom 1983). The right-hand side equation is a variation of MOND, differing in nomenclature, but consistent in practice and results (McGaugh 2012) (McGaugh 2015). Both equations share the same characteristic (universal or critical) acceleration value $a_0 = g† = 1.2 \times 10^{-10}$ ms$^{-2}$ as established from empirical fits demonstrating a tight correlation with low intrinsic scatter. For simplicity, we refer to both above forms of the BFTR as the "MOND" relation as to distinguish it from the proposed 'scaling' version described in this work.

It is believed $a_0$ may represent a fundamental constant of nature, one that is only revealed at galactic scales and in the regime of low gravitational accelerations. From galactic observations and per the above equation, the historically recognized correlation between disk flat rotation velocity and baryonic mass leads to the proposition that a modified gravity law may explain the phenomenology without requiring dark matter (Lelli 2016a). The fit has been so successful in its predictive qualities that it has been deemed a new 'natural law' by its adherents (Lelli 2017).

Within the scaling model, we treat galactic disks as compact finite entities with defined radii equal to the physically observed HI gas disk. In this treatment, disks have flat circular velocities ($V_f$) to the observed edges. Beyond these edges, the rotation velocity breaks into a conventional Keplerian decline following $1/\sqrt{r}$. The combined effect of baryons and their dynamical properties characterize a globalized dynamical mass that is constrained within each galactic disk. This is in contrast to MOND where the flat velocity as measured within the disk is thought to continue to indefinite radii with an asymptotic value that only depends on a galaxy's baryonic mass content $M_{Bar}$ in violation of Newtonian dynamics.

*A Parameterized Model for the BFTR*
In a recent publication, a novel semi-empirical version of the BTFR had been proposed based on the dynamical properties of galactic disks (La Fortune 2019). A subsequent work demonstrated that the BFTR could be reformulated by introducing two supplementary parameters related to the structure and dynamics of individual galaxies based on dimensional and Buckingham/Pi analyses (La Fortune 2020a).

The scaling approach differs from most models by specifying the disk's edge as the physical distance from the galactic center from which measured flat rotation $V_f$ "breaks" into a conventional Keplerian decline and requires use of HI gas radii as provided for most galaxies in the SPARC database. This break-point in the dynamic has been observed in a recent analysis of the Milky Way satellite galaxy population obtained by Gaia (Fritz 2018). Using 3-D satellite velocities, a robust Keplerian decline has been mapped throughout the satellite galaxy sample from 20 to 420 kpc (La Fortune 2020b). This trend is global with the majority of satellites bound within a narrow mass discrepancy window spanning $D=5.9$ to $12.1$ and strongly suggests a point-mass Newtonian potential governing the entire satellite system. We note that this interpretation is in conflict with modified gravity and the $\Lambda$CDM cosmological NFW halo expectations but cannot be argued as a principle dynamical characteristic for the Milky Way and by extension, all rotationally-supported galaxies in the SPARC database.

With above constraint in mind, we introduce two parameters that allow the BTFR to be revised to account for unique structural characteristics specific to each galaxy of interest. These parameters are not new and have been employed extensively in past publications.

The first parameter of interest is galactic mass discrepancy $D=M_{Dyn}/M_{Bar}$ or equivalently $D=V_{Dyn}^2/V_{Bar}^2$. Mass discrepancy is the ratio of the 'missing mass' to that of the baryons as measured at the disk outer edge, $R_D$. $R_D$ is the "terminal radius" where HI gas disk surface density falls to $\approx 1 M_\odot pc^{-2}$ and demarcates the break-point radius separating the disk dynamic from the surrounding Keplerian decline. It is this key feature that differentiates the scaling model from the others. Dynamic mass $M_{Dyn}$ (often loosely termed $M_{Obs}$ or $M_{Total}$) is a function of circular velocity and disk radius per Newton's laws of circular motion: $M_{Dyn}=V_c^2 R_D/G$. For galaxies in the SPARC database, both $V_c$ and $M_{Dyn}$ are provided, giving consistent results using either value within the scaling model.

Baryonic mass $M_{Bar}$ is treated as a central or point mass with velocities following $V_{Bar}=(GM_{Bar}/R_D)^{0.5}$ based on Newton's law of circular motion. From above terms, galactic accelerations are calculated as $g_{Dyn}=g_{Obs}=V_c^2/R_D$ and $g_{Bar}=V_{Bar}^2/R_D$. All pertinent terms related to the MOND and scaling versions of the BTFR are fully quantified with these definitions.

The remaining galactic parameter is dynamic mass surface density which is simply $\mu_{RD}=M_{Dyn}/\pi R_D^2$. This surface density term is also specific to individual galaxies and is easily calculated from data listed in the SPARC database. Unlike the MOND BTFR, we consider this relation dynamically motivated, rather than solely based a galaxy's baryon rest-mass content $M_{Bar}$.

$M_{Dyn}$ is a function of a galaxy's baryonic rest-mass and its included energy and angular momentum as it exists within the disk setting. The BFTR equation below was derived in a previous work based on a semi-empirical fit based on three galactic disk parameters; $M_{Dyn}$, $V_C$ and $R_D$ (La Fortune 2019). This particular equation is considered the nominal or relation of 'central tendency' for rotationally-supported galaxies within the scaling model. For brevity and consistency with previous work, galaxies meeting above parameter criteria are considered to "be on" the scaling [$M_D$-R-V] relation:

$$V_C^4 = M_{Dyn}/256 \quad or \quad \sqrt[4]{M_{Dyn}}/V_C = 4 \leftrightarrow [M_D\text{-R-V}] \text{ relation or "central tendency"}$$

For example, this equation is based on galaxies having mass discrepancies (missing mass) equivalent to the cosmic baryon fraction $f_b$=0.17 (D=5.9) and dynamic mass surface densities equal to $\mu_{MRV}$=67$M_\odot pc^{-2}$. The value of this surface density has special significance in galactic dynamics. It was originally identified as an emergent phenomenon attributed to the self-similarity of dark matter halos radial density profiles (Gentile 2009). Gentile established this characteristic surface density as ‹Σ›$_{0,Dark}$ = 72$M_\odot pc^{-2}$, very close to our assessment and find it is a constant in the denominator in the full scaling BTFR equation. Although these particular values are specified, many other equivalent combinations of D and $\mu_{RD}$ will trace along the [$M_D$-R-V] relation providing a sub-population of galaxies aligning with the central tendency. In a later section, we utilize the right-hand equation and these two parameters to characterize the effective distribution of SPARC galaxies around the central tendency value offering further insight into the true physical nature and behavior of the dynamics governing galactic disks.

A full derivation of the above model is provided in an earlier work as based on a thorough dimensional analysis of the BTFR including a practical example based on the Milky Way Galaxy (La Fortune 2020a). We demonstrate a similarity between the MOND and scaling BFTRs by solving the latter with D=5.9 and $\mu_{MRV}$=67$M_\odot pc^{-2}$. We obtain an equivalent expression to $V_f^4=a_0 GM_{Bar}$ for a galaxy on the [$M_D$-R-V] relation and generate the nominal scaling version of an "acceleration constant" termed $a_S$:

$$V_C^4 = a_s GM_{Bar} \text{ with } a_s = 1.73 \times 10^{-10} \, ms^{-2}$$

It is important to note that this acceleration ($a_S$) is only applicable for galaxies with parameters D and $\mu_{MRV}$ quantified above or equivalent combinations that fall along the [$M_D$-R-V] relation. Since all galaxies are not on the relation, there will be a unique scaling acceleration "$a_S$" value for each galaxy based on its own combination of D and $\mu_{RD}$. This primary feature differentiates MOND from the scaling BTFR, where MOND fixes the acceleration $a_0$ to a resolute constant value rather than a range based on individual galactic properties. We find that $a_S$ is ~50 percent greater than $a_0$ in part due to using specific/measured disk radii (a parameter not employed in the MOND BTFR) to establish galactic dynamics.

Introducing two new parameters D and $\mu_{MRV}$ into the general expression above, the following equation provides a precise relation between $V_C$ and $M_{Bar}$ not previously attainable in the MOND definition. Note there is no use of a universal/characteristic acceleration constant ($a_0$ or g†) nor the gravitational constant G. The scaling BFTR is based on galactic geometry and *dynamical* considerations and can be recast using a "dimensionless" pre-factor rather than a fixed acceleration value:

$$V_C^4 = \frac{D\mu_{RD}}{256\mu_{MRV}} M_{Bar}$$

As opposed to MOND's universal acceleration $a_0$ that applies to all galaxies, $a_S$ is impacted by a galaxy's disk mass discrepancy and dynamic mass surface density. This two-parameter pre-factor offers enhanced flexibility to fit any galaxy in the SPARC database and return accurate modeled circular velocities that precisely match observations.

This approach is more refined than the MOND BTFR, but requires additional information to obtain this improved precision. An advantage of this method over others is that allows a direct link to galactic disk energy and angular momentum content in accordance to the virial theorem (Peebles 1971). This method dispenses with the notion of modified gravity or dark matter velocity support with all components specified by the Peebles spin equation $M_{Dyn}^{5/2}=J\sqrt{E}/\lambda_P G$ where $\lambda_P$ is an observed constant for rotationally-supported disk galaxies, 0.423±0.014 (Marr 2015). In conjunction with the Peebles spin equation and subject to an isothermal constraint $M_{Dyn}=J/\sqrt{2}\lambda_B V_C R_D$, all galactic properties are fully specified provided $\lambda_P=\lambda_B$ (Bullock 2001) (Knebe 2011).

A word of caution must be exercised when comparing these scaling BTFR galactic $M_{Dyn}$ quantities against dark matter virial (halo) mass $M_{Vir}$ or those based on dark matter density contrasts, $M_{200m}$, $M_{200c}$ or splashback mass $M_{SP}$ (Diemer 2020). These ΛCDM virial/total mass estimate do not recognize the dynamic component $M_{Dyn}$ that is driven by baryonic-based disk properties. Within the scaling model, we find $M_{Vir}$ expectedly equivalent to the dark matter halo virial mass thus remaining in agreement between the models in this regard. We like to emphasize virial mass is based on galactic escape velocities ($V_{Esc}=\sqrt{2}V_C$) and are typically ~2X greater than galactic $M_{Dyn}$ data presented herein. $M_{Dyn}$ and $M_{Vir}$ masses are obtained using different methods and therefore should not be used interchangeably.

We see this discrepancy between the definitions in recent ΛCDM cosmological simulations where "dynamic" mass estimates based on rotation velocities fall well short (2-4X) of dark matter halo masses required to obtain comparable rotation curve fits (Posti 2019) (Marasco 2020). Also, cosmological simulations do not specifically define $M_{Dyn}$ as a separate quantity due intrinsic kinetic energy and ordered motion attributed *just* to baryons within the physical disk proper. In many instances, halo total mass and radial density profile are derived without baryonic input or influence with cosmological considerations driving nearly all dark matter halo properties.

To this point, we have presented a classical description of $M_{Dyn}$ employing only baryon mass, energy and angular momentum intrinsic to galactic disks. No dark matter or modified gravity has been introduced to achieve any disk rotation profile. We consider the BTFR a manifestation of the global 'self-similar' properties of galactic disks brought into being by the interplay of Newtonian gravitation, angular momentum and conventionally defined thermodynamic processes that regulate galactic dynamics into equilibrium states in conjunction with their surroundings. As the scaling BTFR is determined solely from baryonic-based disk properties pulled from the SPARC data set, the analysis is classically conventional and quantitative, having no use for 'new physics.'

*The 'Scaling' version of the BTFR as Applied to the SPARC Galaxy Database*
Here we demonstrate the efficacy and practical application of the scaling BTFR using data extracted from the SPARC database. Without appealing to dark matter or modified gravity, we construct our quantitative scaling BTFR model using high quality (Q=1) galaxy data drawn from the Spitzer Photometry and Accurate Rotation Curves (SPARC) database (Lelli 2016b) (Lelli 2020). The SPARC database consists of 173 galaxies rotationally-supported galaxies representing a wide range of morphology, mass, size, and stellar/gas fractions.

From the high-quality data, we eliminated those galaxies having incomplete data resulting in a sample of 81 galaxies used in this analysis. This sample represents galaxies with the cleanest set of observations of the highest accuracy and certainty in their measured values.

This database has been extensively relied upon as evidence for support (or objection) to the theories pertaining to modified gravity theory and dark matter. In addition to $V_C$ and $M_{Bar}$ data (the only variables considered within MOND BTFR), SPARC provides accurate galactic disk radii from detected HI gas which we include to establish the scaling version of the BTFR. We find the SPARC database a powerful resource for advancing/discounting *any* model pertaining to the BTFR.

In Figure 1 below, we provide a template to illustrate the large span of galactic parameters used in this analysis. In the log-log R-V template below, galactic disk radii ($R_D$) are plotted against flat circular velocity ($V_C$). The scatter is not due to measurement error or uncertainty in the estimates, but truly represent the stochastic nature and wide diversity inherent in galactic disks as reported in the SPARC database.

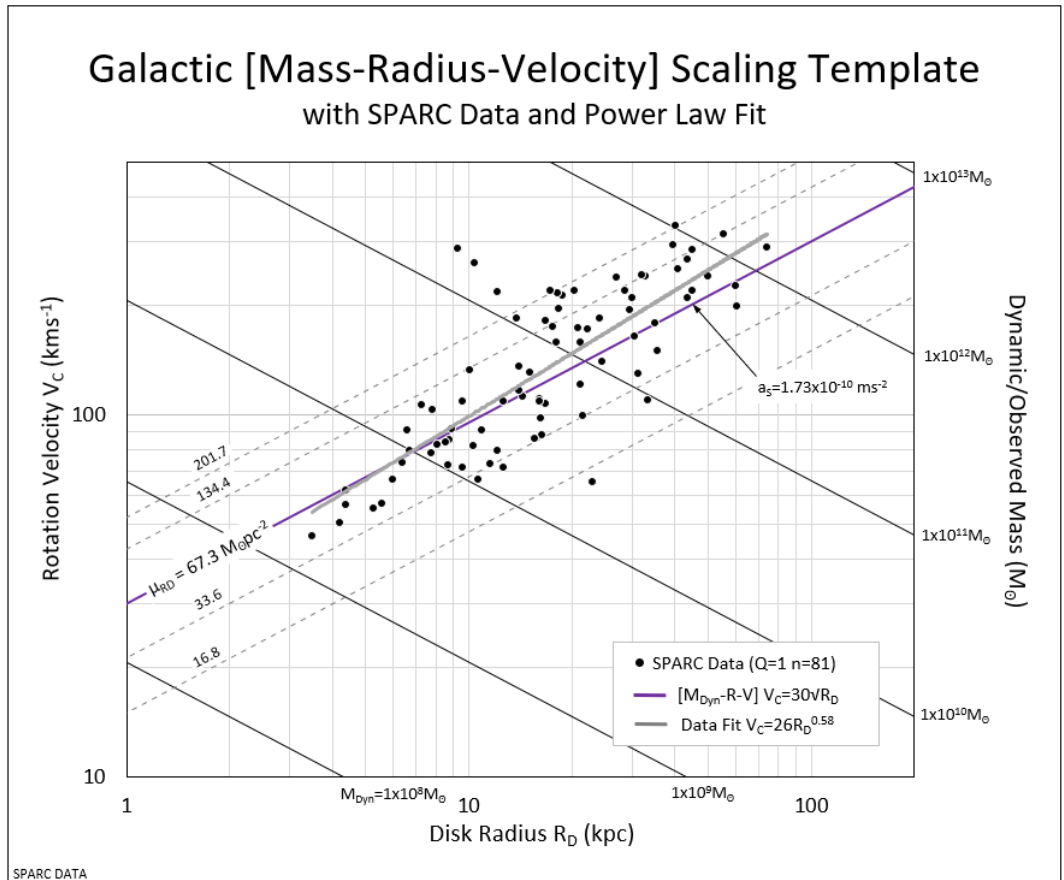

*Figure 1: The Galactic Scaling Template with [$M_D$-R-V] relation (purple solid). The plot plots SPARC galaxies (black points) according to $V_C$ and $R_D$. To guide the eye, $M_{Dyn}$ and $\mu_{RD}$ associated with each galaxy are included. A power law fit (gray solid) illustrates that this sample adheres to the notion of a central tendency (purple solid) per the scaling BFTR equation. Source - (La Fortune 2019)*

Figure 1 depicts an accurate map of the log R-V space inhabited by galaxies attributed to the high-quality SPARC sample. We find galactic structural and kinematic properties to be highly diverse, but still maintaining the strong central tendency alignment with the [$M_D$-R-V] relation (purple solid). The power law fit (gray solid) confirms this agreement between SPARC data and scaling model expectations.

*Characterizing SPARC Galaxies in Mass Discrepancy-Mass Surface Density Space*

In Figure 2, we plot galactic mass discrepancy versus dynamic surface mass density demonstrating the physical relation between these two parameters. The associated MOND $a_0$ and scaling $a_S$ accelerations are included to guide the eye. We observe that the scaling $[M_D$-$R$-$V]$ relation provides a better fit than the MOND acceleration employing $a_0$ as its relation is displaced from the central bulk of the data cluster in this particular parameter space.

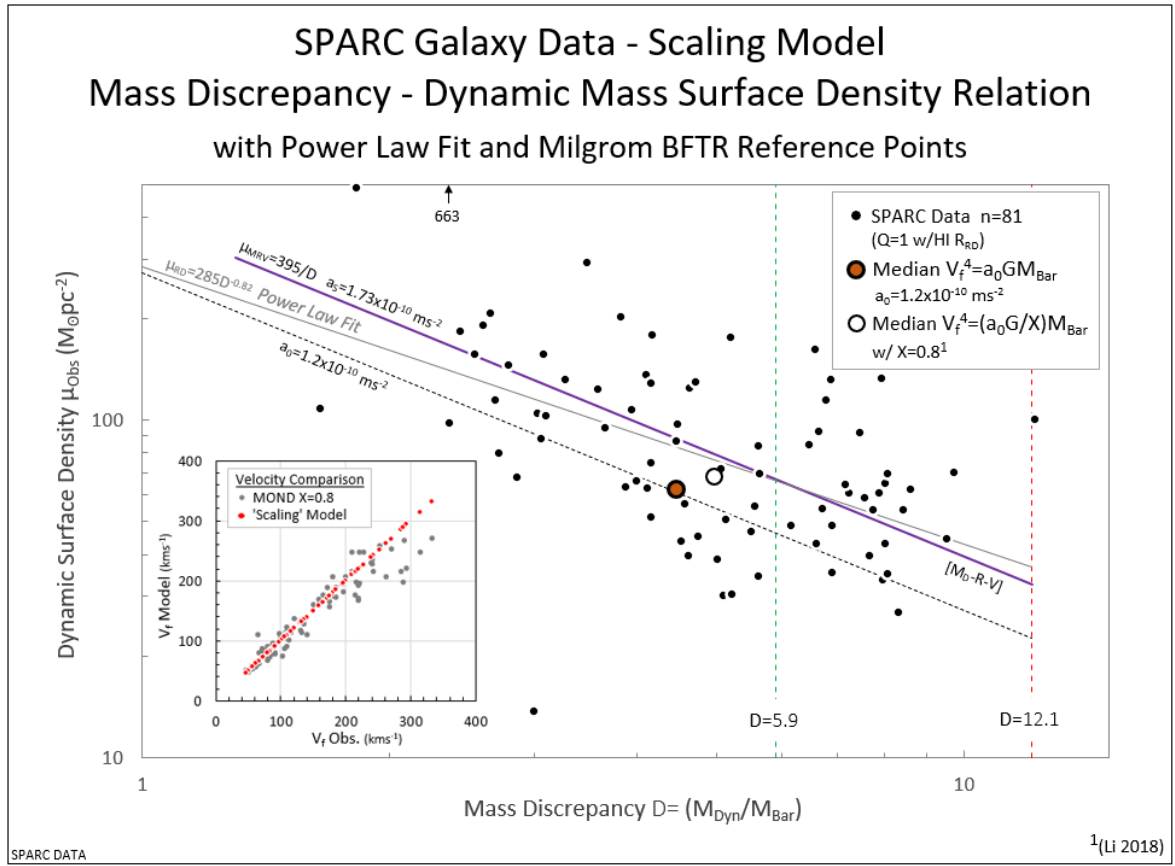

*Figure 2: The SPARC galaxy mass discrepancy-dynamic surface mass density relation. The $[M_D$-$R$-$V]$ relation with $a_S$ constant acceleration (purple solid) is compared against the fiduciary value of $a_0$ (black dash). The MOND galaxy medians for $a_0$ (orange filled black circle) and with the normalization factor (open black circle) within this space is shown for comparison. Per expectation, a power law fit to the data (gray solid) intersects the $[M_D$-$R$-$V]$ relation at $D \approx 5.9$ (green vertical dash) and $\mu_{MRV} \approx 67$ $M_\odot pc^{-2}$. Inset – calculated/modeled velocities against observed $V_C$ for both MOND and scaling BTFRs indicating the improvement when galactic properties are incorporated.*

In the above figure, we include median values attributed to the MOND BTFR for two characteristic accelerations; the first is $a_0$ (orange filled black circle) and the other using a typical normalization factor (open black circle) which, after adjustment, provides better correlations between observed and predicted velocities. To illustrate this effect, we position MOND medians related to their "equivalent" mass discrepancy and surface density solutions fully realizing that they are not parameters inherent in the MOND BTFR equation. We only present these two data points to illustrate quantitative agreement between MOND and scaling solutions.

The normalized acceleration is important as this factor accounts for the finite size of the disk where the terminal disk velocity is normally determined. From a practical standpoint, this factor tightens the correlation between observed and asymptotic velocities shown in Figure 2 inset (Li 2018) (McGaugh 2018). The normalization factor is implemented via the formula $(a_0 G)/X$ with $X=0.8$ being typical. The result of implementing this factor is an increase in $a_0$ ($g\dagger$) to better match measured rotation velocities. Normalizing $a_0$ results in an effective acceleration $a_0/X = 1.5 \times 10^{-10}$ ms$^{-2}$ that is applied equally to all galaxies. We highlight the relevance of this adjustment in terms of modeling parameters and data fits in the galactic mass discrepancy-dynamic surface density space. For example, if $a_0$ was to equal $a_S = 1.73 \times 10^{-10}$ ms$^{-2}$, it would require a normalization factor $X=0.7$ not markedly different than the MOND average typically employed. We find that the normalized $a_0$ value provides a better fit to the SPARC data. This improvement is noted by the visual shift of the MOND median value toward the centroid of the data cluster.

The inset shows the correspondence between observed ($V_{Obs}$) and predicted ($V_{Model}$) circular velocities for both MOND with normalization applied and the parametric scaling relation. Unlike MOND where some scatter is expected (gray points), the scaling model accounts for each galaxy's unique geometry and dynamical behavior resulting in a 1:1 correspondence (red points) between observed and predicted velocities.

The distribution between galaxies in the $D$-$\mu_{RD}$ space represents genuine physical diversity comprising the sample. We find the high-quality data cut demonstrates well-behaved properties with mass discrepancies falling between $D=2$ and $D=10$. This is well inside the limits prescribed by the scaling model where $D=1$ signifies pure Newtonian dynamics and $D=12.1$ which is considered to be the nominal escape velocity relative to baryonic acceleration. We surmise that the generally wide distribution in parameters reflects dependence on intrinsic initial conditions, ongoing secular evolution, and local environmental effects. One must not discount cosmic expansion over time that also can influence galactic final "state properties" and perhaps the value of $a_0$ (Wong 2020).

With mass discrepancy and surface density data available we derive an empirical power law fit that could also be considered a normalization as well for the [$M_D$-R-V] relation. The power law fit show very good alignment with the [$M_D$-R-V] relation and is bounded between the $a_0$ and $a_S$ acceleration isolines. It is interesting to note this empirical power law intersects the MOND characteristic acceleration $a_0$ very close to $D=1$ where Newtonian gravitational dynamics govern and by definition, no 'missing mass' is present or required ($M_{Dyn}=M_{Bar}$). It is not known if this extrapolated fit has true physical significance, but may shed some light on why Milgrom's acceleration value is $a_0=1.2 \times 10^{-10}$ ms$^{-2}$. The extrapolation agrees with the notion that $a_0$ demarcates the transition point below which Newtonian gravitational dynamics are modified.

A recent paper examined the tight correlation in the MOND BFTR using an acceleration constant (Posti 2020). The noted the residuals in the MOND BTFR are uncorrelated with each other bringing into question the role of secondary parameters to improve the fit. They suggest a solution where these secondary parameters could perhaps be anti-correlated, effectively "off-setting" each other. We find this is the case in Figure 2 for the inverse relation between mass discrepancy and dynamic mass surface density. Posti concluded that these parameters can offer insight and understanding pertaining to the dynamical properties of galactic disks while stating that angular momentum plays an important role in achieving a correct relation between $M_{Bar}$ and $V_C$ consistent with the scaling prescription.

In general, the deficiency in the MOND approach is little to no capacity to account for galaxy-to-galaxy diversity only using the characteristic acceleration $a_0$. The advantage of the scaling BTFR is the capability to reproduce the precise observed circular velocity for each galaxy in the SPARC data with a minimum of ad hoc assumptions or corrections.

*The Physical Distribution of the MOND Universal Acceleration - $a_0$ ($g\dagger$)*
Within the 'scaling' model, there is no a priori assumption that a single acceleration analogous to $a_0$ or $g\dagger$ properly explains the $M_{Bar}$-$V_C$ relation. This finding was substantiated in a Bayesian analysis conducted on a large sample of galaxies also obtained from the SPARC database (Rodrigues 2018a). In a follow up study to address concerns in methodology, Marra and Rodrigues repeated their analysis and confirmed the earlier conclusion that $a_0$ is not consistent with a fundamental acceleration scale and that the companion RAR should be considered an emergent phenomenon and not connected to a new natural law (Marra 2020).

We illustrate the distribution of scaling derived accelerations in a format first presented by Rodrigues in Figure 3 below. Individual galactic acceleration values obtained using the scaling model are ordered in ascending by galaxy (black points) with a comparison to Marra's acceleration distribution (gray solid) and best fit (gray dash). We include the constant MOND acceleration constant $a_0$ (red dash) for comparison.

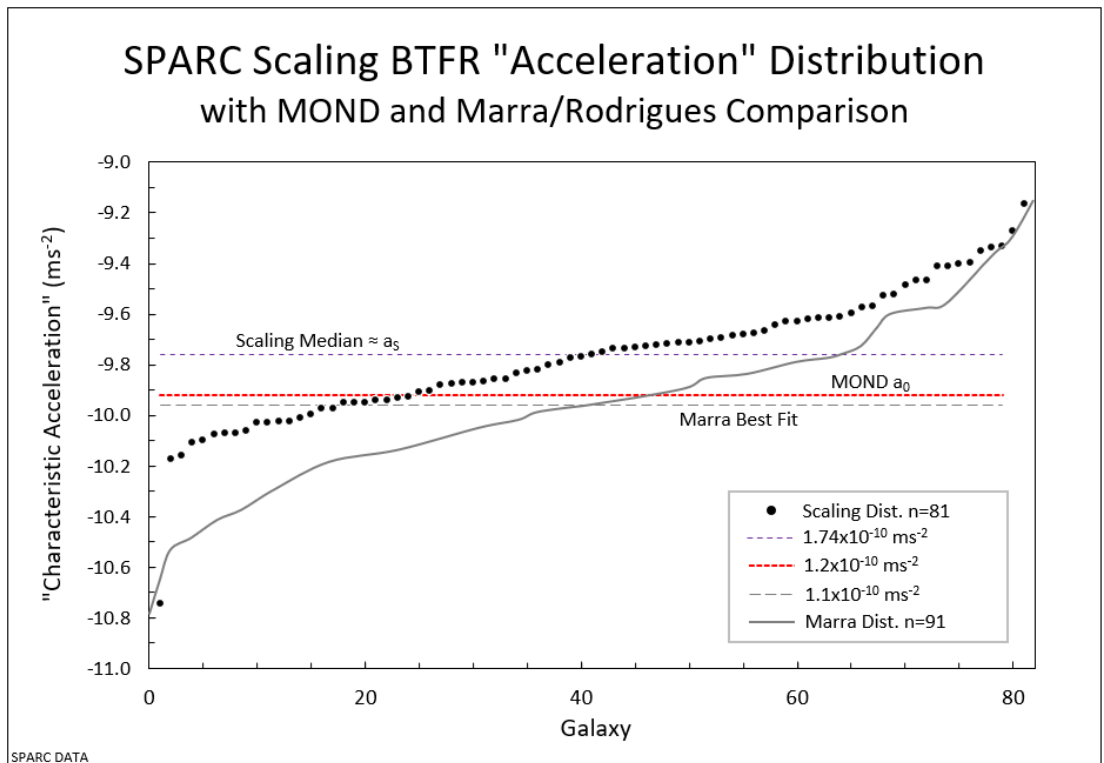

*Figure 3: Characteristic accelerations for high-quality data culled from the SPARC galaxy database. The distribution of individual galaxy scaling accelerations (black data points) agrees with previous Bayesian analysis by Marra/Rodrigues. A modest offset in the magnitude of the "best fit" acceleration values (gray solid) is shown with the scaling values demonstrating a relatively flat profile throughout its range. MOND $a_0$ is provided for comparison (red dash) indicating a close fit with Marra's best fit. Source - (Marra 2020)*

We find the overall range in accelerations compatible with the Bayesian analysis, with the upward shift in the average value being the primary difference. The scaling distribution is flatter than Marra which can give the impression of a nearly constant value for the majority of the sample, although both sets of results span essentially the same range - roughly one dex. Conversely, MOND's calculated intrinsic scatter of $a_0$ is a very narrow ~0.6 dex found using an alternate approach (Li 2018).

Like the more advanced analyses, our results argue against a universal for fundamental acceleration that is a cornerstone of MONDian theory. Rather than just reporting the fits with low explanatory power, the scaling methodology provides a physical foundation and prediction capability for individualized galactic accelerations. Moreover, as the measured accelerations always fall on the RAR at some point along the curve, it is evident that the RAR consists of a large family of relations, each one galaxy specific (McGaugh 2016).

From a first principles perspective, it would seem natural that global disk acceleration ($a_S$) should be dependent on mass discrepancy and dynamic mass surface area and that they should also be anti-correlated. Unfortunately, this 'offsetting' behavior between these two salient parameters has given the false impression that $a_0$ is a fundamental property associated with rotationally-supported disk galaxies. Despite strong evidence against $a_0$ being fundamental, there has been ongoing spirited debate over the intervening years without a final resolution in hand (McGaugh 2018) (Rodrigues 2018b) (Chang 2019) (Cameron 2020) (Rodrigues 2020).

*The Distribution Around the Scaling BFTR 'Central Tendency' Value*
We can glean additional information by plotting the ratio of galactic dynamic mass to circular velocity for each galaxy in the SPARC dataset through an analysis of the distribution of galactic properties around the scaling central tendency $M_{Dyn}^{0.25}/V_C=4$. For this analysis, we create a Cumulative Distribution Function (CDF) shown in Figure 4 and compare it to a Gaussian distribution inferring a possible thermodynamic origin and equilibrium states for observed scatter in the SPARC galaxy sample.

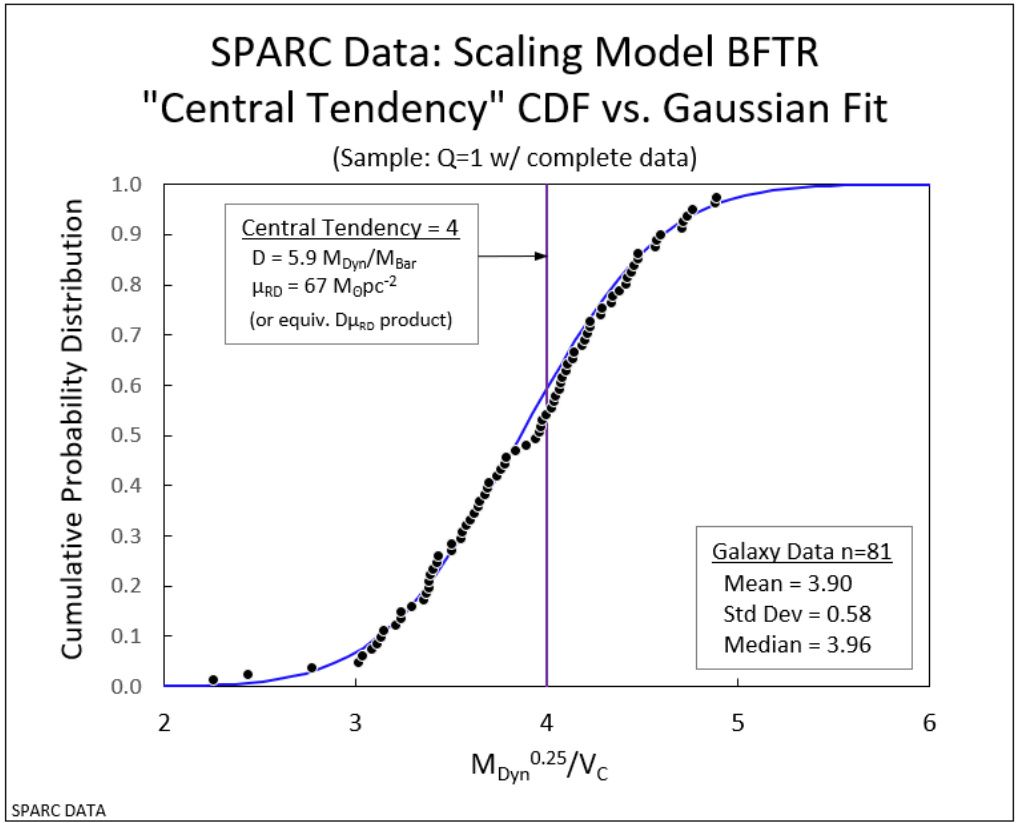

*Figure 4: The SPARC sample Cumulative Distribution Function (black points) compared to the scaling BTFR 'central tendency' value =4 (purple solid), i.e., the [$M_D$-R-V] relation. The data is fit to a Gaussian CDF (blue solid) with the parameters identified in the key. This distribution infers a random distribution around the mean which is a hallmark of statistical variability and infers global thermodynamically regulated processes governing each galaxy.*

Figure 4 reveals SPARC galaxies distribute around the central tendency with its CDF well described by a Gaussian distribution. We find the mean ($\mu$=3.9) falls slightly below the central value 4 (vertical purple) which is consistent with in the power law fit shown in Figure 2 being slightly shallower than the [$M_D$-R-V] scaling relation expectation.

We can attribute this Gaussian distribution to the largely stochastic/random variability due to thermodynamic processes driven to different configurations of long-lived equilibrium states. It is a signature that galactic dynamics are governed by classically described physical processes and pure Newtonian dynamics with no need to invoke dark matter or modified gravity. These new relations and constraints provide a 'pressure-test' for any galactic or cosmological model, regardless of motivation.

*Summary*
A two-parameter model for the BTFR has been constructed using data provided in the SPARC galaxy database. By introducing the simple notion of flat rotation to disk edges followed by a conventional Keplerian velocity decline beyond, two important 'galaxy specific' properties can be defined; mass discrepancy and dynamic mass surface density. These parameters, in conjunction with an alternative, dynamically-based BTFR formula provides precise correspondence between baryonic mass and circular velocity on a galaxy-by-galaxy basis. This level of precision has not been attainable with current expressions of the BFTR utilizing a characteristic acceleration constant.

*Acknowledgements*

We thank those assisting in the development of this proposal. We thank arXiv for its preprint platform and service.